\documentclass[conference]{IEEEtran}
\IEEEoverridecommandlockouts

\usepackage{times}
\usepackage[cmex10]{amsmath}
\usepackage{amssymb}
\usepackage{epsfig,verbatim}
\usepackage{algorithm,algpseudocode}
\usepackage{hhline}
\usepackage{algpseudocode}
\usepackage{booktabs}
\usepackage{multirow}
\usepackage{graphicx, subfigure}
\usepackage{cite}
\long\def\comment#1{}

\newfont{\bbb}{msbm10 scaled 700}

\newfont{\bb}{msbm10 scaled 1100}
\newcommand{\CC}{\mbox{\bb C}}
\newcommand{\PP}{\mbox{\bb P}}
\newcommand{\RR}{\mbox{\bb R}}

\newcommand{\av}{{\bf a}}

\newcommand{\cv}{{\bf c}}

\newcommand{\hv}{{\bf h}}

\newcommand{\qv}{{\bf q}}
\newcommand{\rv}{{\bf r}}

\newcommand{\wv}{{\bf w}}

\newcommand{\xv}{{\bf x}}

\newcommand{\zv}{{\bf z}}

\newcommand{\Hm}{{\bf H}}

\newcommand{\Cc}{{\cal C}}
\newcommand{\Dc}{{\cal D}}

\newcommand{\Lc}{{\cal L}}
\newcommand{\Mc}{{\cal M}}
\newcommand{\Nc}{{\cal N}}

\newcommand{\Sc}{{\cal S}}

\newcommand{\Uc}{{\cal U}}

\newcommand{\epsilonv}{\hbox{\boldmath$\epsilon$}}

\newcommand{\thetav}{\hbox{\boldmath$\theta$}}

\newcommand{\sign}{{\hbox{sign}}}

\newcommand{\SNR}{{\sf SNR}}

\renewcommand{\Re}{{\rm Re}}
\renewcommand{\Im}{{\rm Im}}
\newcommand{\eqdef}{\stackrel{\Delta}{=}}

\newcommand{\transp}{{\sf T}}

\newtheorem{lemma}{Lemma}

\newcommand{\argmax}{\operatornamewithlimits{argmax}}

\title{Semi-Supervised Learning Detector for MU-MIMO Systems with One-bit ADCs}

\author{
\IEEEauthorblockN{
              Seonho Kim and Song-Nam Hong}
\IEEEauthorblockA{Ajou University, Suwon, Korea,\\
              email: \{kimsh1005  and snhong\}@ajou.ac.kr}
}

\begin{document}
\maketitle


\begin{abstract}  

We study an uplink multiuser multiple-input multiple-output (MU-MIMO) system with one-bit analog-to-digital converters (ADCs). For such system, a supervised-learning (SL) detector has been recently proposed by modeling a non-linear end-to-end system function into a parameterized Bernoulli-like model. Despite its attractive performance, the SL detector requires a large amount of labeled data (i.e., pilot signals) to estimate the parameters of the underlying model accurately. This is because the amount of the parameters grows exponentially with the number of users. To overcome this drawback, we propose a semi-supervised learning (SSL) detector where both pilot signals (i.e., labeled data) and some part of data signals (i.e., unlabeled data) are used to estimate the parameters via expectation-maximization (EM) algorithm. Via simulation results, we demonstrate that the proposed SSL detector can achieve the performance of the existing SL detector with significantly lower pilot-overhead.


\end{abstract}


\section{Introduction}


Massive multiple-input multiple-output (MIMO) is a promising technology for beyond 5G cellular systems where a large number of antennas at the BS is used to improve the capacity and energy-efficiency  \cite{Lu}. In contrast, it can cause the hardware cost and the radio-frequency (RF) circuit power consumption to increase significantly \cite{Yang}. Especially, a high-resolution analog-to-digital converter (ADC) is the major problem as the power consumption of an ADC increases exponentially with the number of quantization bits and linearly with the baseband bandwidth\cite{Mezghani-2011}. To overcome this challenge, the use of low-resolution ADCs (e.g., 1$\sim$3 bits) for massive MIMO systems has received increasing attention over the past years. The one-bit ADC is particularly attractive as there is no need for an automatic gain controller, which reduces the hardware complexity significantly\cite{Hoyos}. In this case,  simple zero-threshold comparators quantize the in-phase and quadrature components of the continuous-valued received signals separately. Although low-resolution ADCs provides the advantages, it gives rise to numerous technical challenges in channel estimation and MIMO detections.


For uplink MU-MIMO systems with one-bit ADCs, numerous channel estimation methods were developed as least-square (LS) based method \cite{Risi}, maximum likelihood (ML) method\cite{Choi}, zero-forcing (ZF) type method \cite{Choi} and Bussgang decomposition based method \cite{Li}. Also, regarding MIMO detections, the optimal ML detection was developed in \cite{Choi}, and the low complexity methods were presented in \cite{Mollen, Mollen2}. Inspired by coding theory, the MIMO detection problems have been reconstructed as an equivalent coding problem \cite{Hong}. Using the resulting model, a weighted minimum distance (wMD) decoding (i.e., an alternative expression of the ML detector) was presented. Very recently, supervised-learning (SL) detectors were proposed in \cite{Pohang_Sup,Sup,Reinfor} for the considered communication system with one-bit quantized signals. Especially, in our prior work \cite{Sup}, we proposed the generative model, called Bernoulli-like model, by considering the traits of one-bit quantized signals. Despite its attractive performance, the SL detector in \cite{Sup} requires a large amount of pilot overhead to estimate the model parameters accurately. Thus, it is necessary to reduce a pilot overhead so that the SL detector will be used in practical systems.



\begin{figure}
\centerline{\includegraphics[width=9cm]{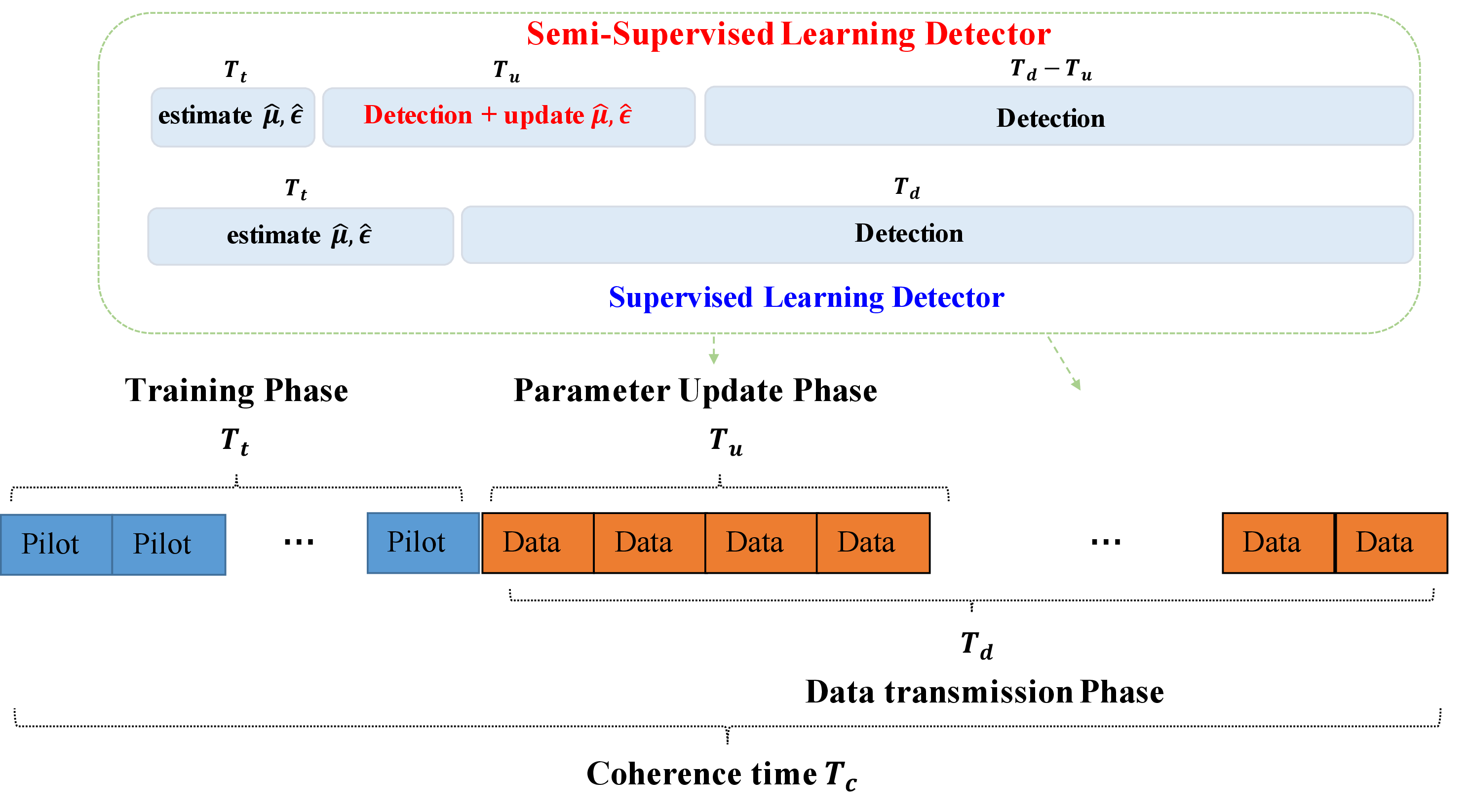}}
\caption{Illustration of training, parameter-update, and data transmission phases within a coherence time.}
\label{phase}
\end{figure}

In this paper, we study an uplink MU-MIMO system with one-bit ADCs where $K$ users with single-transmit antenna communicate with one BS with $N_{\rm r}$ receive antennas. Also, it is assumed that the BS is not aware of a channel state information (CSI) as in practical communication systems, and needs to estimate it using pilot signals during training phase (see Fig.~\ref{phase}). A block-fading channel is assumed in which the channel is static during the coherence time $T_c$ and changes independently in block-to-block. We assign the first $T_{t} < T_c$ time slots to the channel training phase and the remaining $ T_ {d} = T_ { c} - T_{t} $ time slots are dedicated to the data transmission phase as shown in Fig. \ref{phase}. Inspired by semi-supervised learning\cite{zhu2009introduction}, for such system, we propose a semi-supervised learning (SSL) detector which can significantly reduce the pilot-overhead of the existing SL detector in \cite{Sup}. The main idea of the proposed SSL detector is that it uses both pilot signals (i.e., labeled data) and some part of data signals (i.e., unlabeled) data to estimate the parameters of the underlying Bernoulli-like model via an efficient expectation-maximization (EM) algorithm. Via simulation results, we demonstrate that the proposed SSL detector can achieve the same performance of the SL detector with a significantly reduced pilot-overhead (e.g., $50\%$ overhead reduction). 

This paper is organized as follows. In Section \ref{sec:Preliminaries}, we describe an uplink MU-MIMO system with one-bit ADCs and equivalent parallel binary discrete memoryless channels in a coding-theoretic viewpoint. In Section \ref{sec:Supervised_Learning}, we briefly review a SL detector for the considered system. In Section \ref{sec:SSL}, we propose a novel SSL detector with parameter update rules which are built on EM algorithm. Section \ref{simulation} provides the simulation results to verify the superiority of the proposed SSL detector. Finally, conclusion is provided in Section \ref{conclusion}.

 {\bf Notation:} Lower and upper boldface letters represent column vectors and matrices, respectively. Let $[a:b]\eqdef\{a,a+1,\ldots,b\}$ for any integers $a$ and $b>a$, and when $a=1$, it can be further shortened as $[b]$.  For any $k \in [0:K-1]$, we let $g(k)=[b_0,b_1,\ldots,b_{K-1}]^{\transp}$ represent the $m$-ary expansion of $k$ where 
$k=b_0m^0+\cdots+b_{K-1}m^{K-1}$ for $b_i \in [0:m-1]$.
 We also let $g^{-1}(\cdot)$ denote its inverse function. For a vector, $g(\cdot)$ is applied element-wise. Likewise, if a scalar function is applied to a vector, it will be performed element-wise. $\Re(\av)$ and $\Im(\av)$ represent the real and complex part of a complex vector $\av$, respectively.

\section{Preliminaries}\label{sec:Preliminaries}

In this section, we describe the system model and define an equivalent $N$ parallel binary discrete memoryless channels (DMCs).

\subsection{System model} \label{system}

We consider a single-cell uplink MU-MIMO system in which $K$ users with a single-antenna communicate with one BS with an array of $N_{\rm r} > K$ antennas. We denote $w_k \in \mathcal{W}=[0:m-1]$ as the user $k$'s message for $k \in [K]$, each of which contains $\log{m}$ information bits. Also Let $m$-ary constellation set by $\Sc=\{s_0,...,s_{m-1}\}$ with power constraint as 
\begin{equation}
\frac{1}{m}\sum_{i=0}^{m-1} |s_i|^2 = \SNR.
\end{equation} At time slot $t$, the user $k$ transmits the symbol ${\tilde x}_k[t]$ as 
\begin{equation}
\tilde{x}_k[t] = \Mc(w_k[t]) \in \Sc,
\end{equation} where  $\Mc:\mathcal{W}\rightarrow \Sc$ denotes a modulation function.
When all the $K$ users transmit the symbols ${\tilde \xv[t]}=[\tilde{x}_1[t],\ldots,\tilde{x}_K[t]]^{\transp}$, the BS receives the discrete-time complex-valued baseband signal vector ${\bf \tilde {r}[t]}\in\mathbb{C}^{N_{\rm r}}$,  given by
\begin{equation}
{\bf \tilde{r}}[t] = {\bf \bar H}{\bf \tilde{x}}[t] +{\bf \tilde{z}}[t], \label{eq:system_complex}
\end{equation} where ${\bf \tilde H} \in \CC^{N_{\rm r} \times K}$ is the channel matrix between the BS and the $K$ users, for example, the $i$-th row of ${\bf \tilde H}$ is the channel vector between the $i$-th receiver antenna at the BS and the $K$ users. Also, ${\bf \tilde{z}}[t]=[{\tilde z}_1[t],\ldots,{\tilde z}_{N_{\rm r}}[t]]^{\transp}\in\mathbb{C}^{N_{\rm r}}$ denotes the noise vector whose elements are distributed as circularly symmetric complex Gaussian random variables with zero-mean and unit-variance, i.e., ${{\tilde z}_i}[t] \sim \Cc\Nc(0,1)$. 

In the MIMO system with one-bit ADCs, each receiver antenna of the BS is equipped with RF chain followed by two one-bit ADCs that are applied to each real and imaginary part respectively.  We define $\mbox{sign}(\cdot): \RR \rightarrow \{-1,1\}$ as the one-bit ADC quantizer function with $\hat{r}[t]=\mbox{sign}(\tilde{r}[t]) = 1$ if $\tilde{r}[t] \geq 0$, and $\hat{r}[t]=-1$, otherwise.  Then, the BS receives the quantized output vector as $\hat{\rv}_{\rm R}[t] = \mbox{sign}({\rm Re}({\bf \tilde r}[t]))$ and $\hat{\rv}_{\rm I}[t] =\mbox{sign}({\rm Im}({\bf \tilde r}[t]))$. For the ease of representation, we rewrite the complex input-output relationship in \eqref{eq:system_complex} into the equivalent real representation as
\begin{equation}
\rv[t] = \mbox{sign}\left(\Hm\xv(\wv[t])+\zv[t]\right), \label{eq:obs1}
\end{equation}
where $\rv[t]=[\hat{\rv}_{\rm R}^{\transp}[t],\hat{\rv}_{\rm I}^{\transp}[t]]^{\transp}$, $\xv(\wv[t])=[\Re(\tilde{\xv}[t])^{\transp},\Im(\tilde{\xv}[t])^{\transp}]^{\transp}$, $\zv[t]=[\Re(\tilde{\zv[t]})^{\transp},\Im(\tilde{\zv}[t])^{\transp}]^{\transp}\in\mathbb{R}^{N}$, and 
\begin{equation*}
\Hm = \left[ {\begin{array}{cc}
   \Re({\bf \tilde H}) & -\Im({\bf \tilde H}) \\      
   \Im({\bf \tilde H}) & \Re({\bf \tilde H})\\
 \end{array} } \right] \in\mathbb{R}^{N\times 2K},
 \end{equation*}
where $N=2N_{\rm r}$. This real system representation will be used in the sequel.

\subsection{Equivalent N parallel B-DMCs} \label{system}
In \cite{Hong}, it was shown that a real system representation \eqref{eq:obs1} can be transformed into an equivalent $N$ parallel B-DMCs via a coding-theoretic viewpoint. In the resulting $N$ parallel B-DMCs, the channel input/output and the channel transition probabilities are defined as follows.

\begin{figure}
\centerline{\includegraphics[width=9cm]{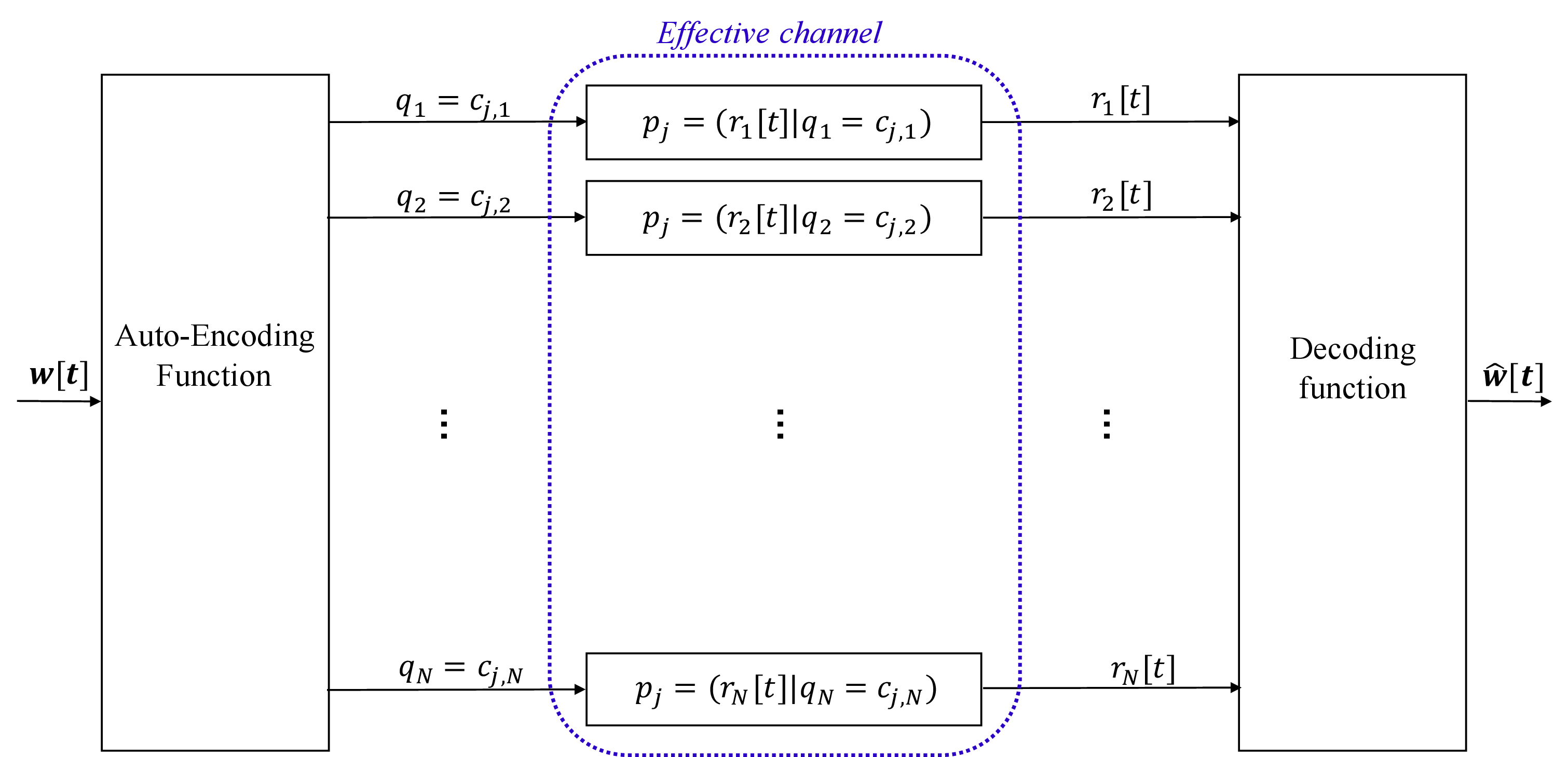}}
\caption{Description of the equivalent $N$ parallel B-DMCs.}
\label{System}
\end{figure}

{\bf Auto-encoding function:} Given $\Hm$, we can create a spatial-domain code $\Cc=[\cv_0,\ldots,\cv_{m^K-1}]$, each of which is given by
\begin{equation}
\cv_j= \left[\mbox{sign}\left(\hv_{1}^{\transp}\xv(g(j))\right),\ldots, \mbox{sign}\left(\hv_{N}^{\transp}\xv(g(j))\right)\right]^{\transp}\label{codecon}
\end{equation} where note that each codeword of $\Cc$ can be considered as a noiseless channel output in \eqref{eq:obs1}. In Fig.~\ref{System}, the channel input $\qv$ of the equivalent channel is determined by the auto-encoding function $f(\cdot)$ such as 
\begin{equation}
\qv=f(\wv,\Hm)=\cv_j,
\end{equation} for $j=g^{-1}(\wv)\in [0 : m^K-1]$.
\vspace{0.1cm}

{\bf Effective channel:} As shown in Fig.~\ref{System}, the effective channel consists of the $N$ parallel BSCs with the channel input $\qv$ and the channel output $\rv$. This channel is specified by the following channel transition probabilities:  For the $n$-th BSC, the transition probability, depending on user's message $\wv=g(j)$ and the corresponding codeword $\cv_j$, are defined as
\begin{equation}\label{eq:epsilon}
\PP(r_n[t]|q_n=c_{j, n})=\begin{cases} 
      \epsilon_{j,n} & \text{if}~ r_n[t]\neq c_{j,n} \\
      1-\epsilon_{j,n} & \text{if}~ r_nl[t]=c_{j,n}
   \end{cases}
\end{equation} where the error-probability of the $n$-th BSC is computed as
\begin{equation}
\epsilon_{j,n} \eqdef Q\left(|\hv_{n}^{\transp}\xv(g(j))|\right), \label{eq:epsilon_1}
\end{equation} where $Q(x) = \frac{1}{2\pi}\int_{x}^{\infty} \exp\left(-u^2/2\right) du$.

The purpose of this paper is to design a decoding function in Fig.~\ref{System} which decodes $\hat{\wv}[t]$ from an observation $\rv[t]$, by leveraging the equivalent effective channel (i.e., the channel transition probabilities in (\ref{eq:epsilon})). We remark that the parameters of the transition probabilities are not known a priori and should be estimated with pilot signals during the training phase.

\section{The Overview of SL Detector}\label{sec:Supervised_Learning}

In this section, we briefly review the supervised-learning (SL) detector proposed in \cite{Sup} with the assumption that a channel matrix $\Hm$ is not known. In the SL detector, thus, we need to estimate the parameters $\Cc$ and $\epsilon_{j,\ell}$ using pilot signals as in parameterized supervised learnings. From (\ref{eq:epsilon}), we can define the generative model of $\rv[t]$, named {\em Bernoulli-like} model, which are fully described by the parameter vector $\thetav=[\thetav_0,\ldots,\thetav_{m^K-1}]$ where $\thetav_j = [\cv_j,\epsilonv_j]$, such as 
\begin{align}
p(\rv[t] |j,\thetav_{j})&\eqdef \PP(\rv[t]|g^{-1}(\wv[t])=j, \thetav_j))\nonumber\\
& =\prod_{n=1: r_n[t]\neq c_{j,n}}^{N}\epsilon_{j,n}\prod_{n=1: r_n[t] = c_{j,n}}^{N}(1-\epsilon_{j,n}) \label{eq:BLmodel}
\end{align} for $j \in [0:m^K-1]$.
We remark that each class $j$ has its own probability distribution parameterized by $\thetav_j=[\cv_j, \epsilonv_j]$.

The SL detector in \cite{Sup} performs with the following two-phase during each coherence time $T_{c}$.

{\bf Parameter Estimation:} In this phase, the parameter vector $\thetav$ is estimated using  $T_t$ pilot signals. We first obtain the labeled data $\Lc$ such as
\begin{equation}\label{Pilots}
\Lc=\{(\rv[1], 0),\ldots,(\rv[T],0),\ldots,(\rv[T_t],m^K-1)\},
\end{equation} where $(\rv[t], j_t)$ represents the pilot signal corresponding to the label $j_t$. Since $T$ pilot signals are transmitted for each codeword, the overall pilot-overhead is equal to $T_{\rm t}=T\cdot {m^K}$. Also, for $t \in [T_t]$, the labels are determined as
\begin{equation}\label{eq:label}
j_t \eqdef \lfloor (t-1)/T \rfloor \in [0:m^K-1],
\end{equation} $\lfloor \cdot \rfloor$ denotes the floor function. In \cite{Sup}, from the labeled data $\Lc$, the parameter vector $\thetav$ is determined via the optimal maximum-likelihood (ML) estimation as
\begin{align}
\hat{c}_{j, n}&=\mbox{sign}\left(\sum_{t=j\cdot{T}+1}^{(j+1)\cdot{T}}r_{n}[t]\right) \label{eq:opt_mu}\\
\hat{\epsilon}_{j,n}&=\frac{1}{T}{\sum_{t=j\cdot{T}+1}^{(j+1)\cdot{T}} {\bf 1}_{\{\hat{\cv}_{j,n}\neq r_{n}[t]\}}}\label{eq:est_epsilon}
\end{align} for $n\in [N]$ and $j \in [0:m^K-1]$.
\vspace{0.2cm}

{\bf Data Detection:} From the Bernoulli-like model parameterize by (\ref{eq:opt_mu}) and (\ref{eq:est_epsilon}), the ML detection performs as
\begin{equation}
\hat{j}=\argmax_{j\in [0:m^K-1]}{p(\rv[t] |j,\thetav_{j})}.
\label{eq:detects} 
\end{equation}

\section{The Proposed SSL Detector}\label{sec:SSL}

Despite its superior performance, the SL detector proposed in \cite{Sup} suffers from the heavy pilot-overhead because a larger number of pilot signals are required so that an empirical transition probability in (\ref{eq:est_epsilon}) is close to the true transition probability in  \eqref{eq:epsilon_1}. Moreover, this overhead becomes larger as the number of users $K$ increases, because the number of parameters to be estimated increases exponentially with the $K$ (see (\ref{eq:opt_mu}) and (\ref{eq:est_epsilon})). To address the above problem, we propose a semi-supervised learning (SSL) detector in which the parameter vector $\thetav$ is estimated by leveraging both data signals (i.e., unlabeled data $\Uc$) and pilot signals (i.e., labeled data $\Lc$). Here, the unlabeled data $\Uc$ is collected during $T_{u}$ time slots (see Fig.~\ref{phase}) such as
\begin{equation}\label{Pilots}
\Uc=\{\rv[T_t+1],\rv[T_t+2],\ldots,\rv[T_t+T_u]\}.
\end{equation}  Also, we let $\Dc = \Lc \cup \Uc$ denote the observed data to be used for parameter-estimation in the proposed SSL detector.

\begin{figure*}
\centerline{\includegraphics[width=17cm,height=7cm]{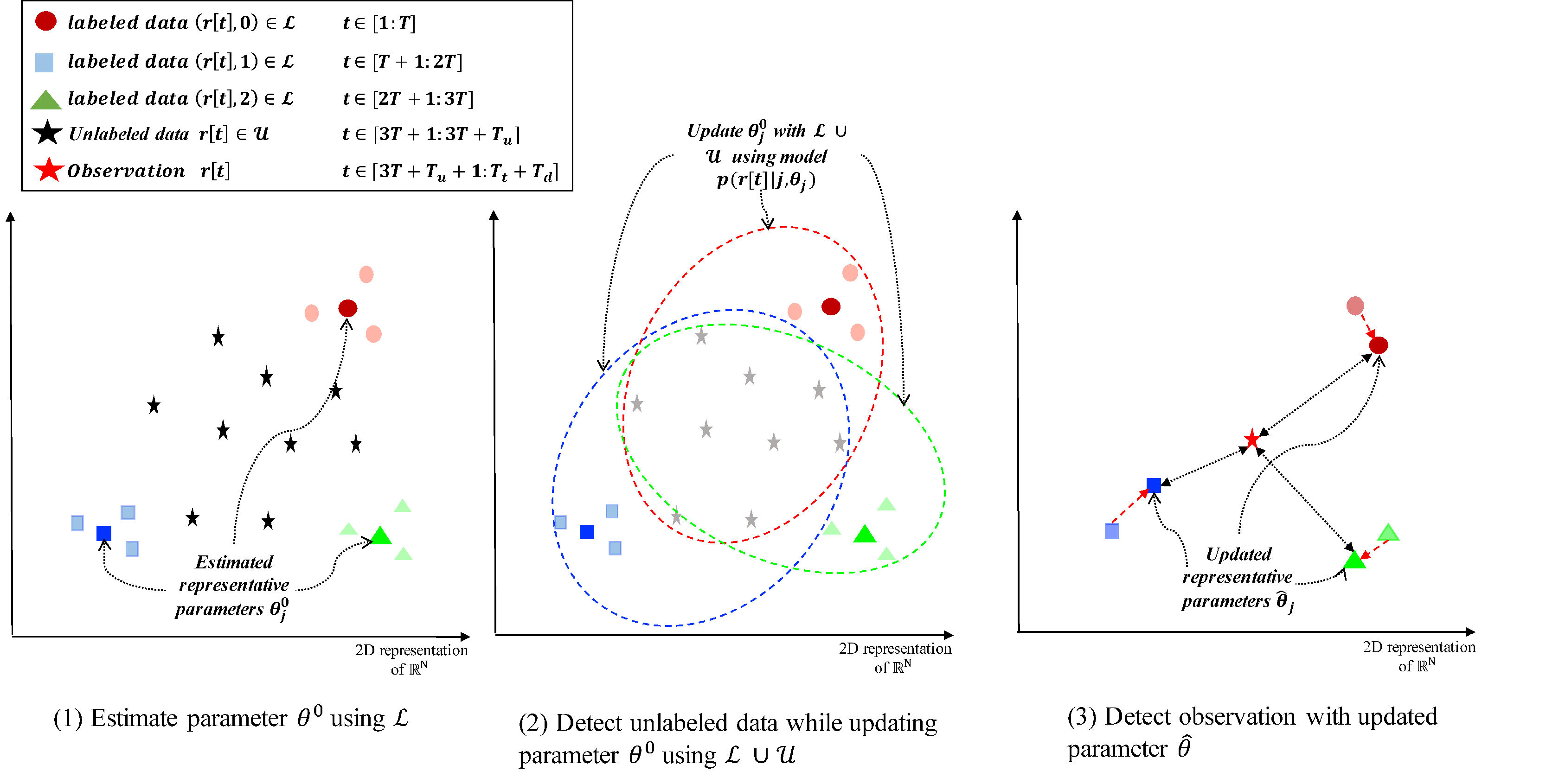}}\vspace{-0.3cm}
\caption{Illustration of overall procedures of the proposed SSL detector when $T=3$ and $j\in[0:2]$ }
\label{SSL concept}
\end{figure*}

{\bf Parameter Estimation:} In this phase, the parameter vector $\thetav=[\thetav_0,\ldots,\thetav_{m^K-1}]$ is updated from the given data $\Dc$ so that the conditional probabilities of the observations (i.e., the received binary signals) are maximized. This ML estimation is mathematically formulated as 
\begin{align}\label{eq:ML}
\hat{\thetav} = \argmax_{\thetav}{\log \PP(\Dc|\thetav)}.
\end{align} Note that from the Bernoulli-like model, we know the probability distribution $p(\rv[t]|j,\thetav_j)$ defined in (\ref{eq:BLmodel}) for the given parameter $\thetav_j$, which will be used in the below. Also, the labels of the labeled data ate given as $\{j_t=\lfloor (t-1)/T\rfloor: t\in[T_t]\}$ in (\ref{eq:label}).

For any fixed parameter $\thetav$, the objective function in (\ref{eq:ML}) is represented as
\begin{align}
&\log{\PP(\Dc|\thetav)}\nonumber\\
&=\log{\prod_{t=1}^{T_t}\PP(\rv[t],g^{-1}(\wv[t])= j_{t} |\thetav_{j_{t} }) \prod_{t=T_t+1}^{T_t+T_u} \PP(\rv[t]|\thetav)}\nonumber\\
&=\sum_{t=1}^{T_t} \log{\PP(j_{t}  |\thetav_{j_{t} }) p(\rv[t]|j_{t} ,\thetav_{j_{t} })}\nonumber\\
&\;\;\;\;\;\;\;\;\;\;\;\;\;\;\;\;\;\;\;+\sum_{t=T_t+1}^{T_t+T_u} {\log\left({\sum_{j=0}^{m^K-1} p(\rv[t], j|\thetav_{j})}\right)},\label{eq:likelihood}
\end{align}  where recall that $p(\rv[t] |j,\thetav_{j})$ is defined in (\ref{eq:BLmodel}), and $\PP(j_{t}  |\thetav_{j_{t}}) = 1/m^K$ since the users' messages are assumed to be generated uniformly and randomly. Definitely, the above objective function is non-convex especially due to the second-term caused by the unlabeled data and thus, the optimization problem in (\ref{eq:ML}) is too complex to be solved. We thus solve it using {\em Expectation-Maximization} (EM) algorithm\cite{dempster1977maximum}.

The EM algorithm consists of the following two steps, named expectation-step (E-step) and maximization-step (M-step), respectively: Given the up-to-date parameter vector $\thetav^i$, it finds the updated parameter vector $\thetav^{i+1}$.

{\em E-step:} In this step, we compute the following probability distribution using the latest parameter vector $\thetav^i$:
\begin{align}\label{eq:expect}
\gamma_{j}[t] &\eqdef \PP(g^{-1}(\wv[t])=j|\rv[t],\thetav_{j}^{i}).
\end{align} This is specified by considering the difference of the labeled and unlabeled data as follows:
\begin{itemize}
\item {\em (Labeled Data)} For $t \in [T_t]$ and $j \in [0:m^K-1]$, 
\begin{equation}
\gamma_{j}[t]={\bf 1}_{\{j=j_{t} \}}.
\end{equation}
\item {\em (Unlabeled Data)} For $t \in [T_t+1:T_t+T_u]$ and  $j \in [0:m^K-1]$,
\begin{equation}
\gamma_{j}[t]=\frac{p(\rv[t]| j,\thetav_j^{i})}{\sum_{j=0}^{m^K-1}{p(\rv[t]|j,\thetav_{j}^{i})}}.
\end{equation}
\end{itemize}

{\em M-step:} In this step, we find an updated parameter vector $\thetav^{i+1}$ using the $\gamma_j [t]$ in the above  as follows:
\begin{equation}\label{eq:obj_EM}
\thetav^{i+1}=\argmax_{\thetav}{\psi(\thetav|\thetav^i)},
\end{equation} where the objective function is defined as
\begin{align}
&\psi(\thetav|\thetav^i)\nonumber\\
&\;\;\;\;\;\eqdef{\sum_{t=1}^{T_t+T_u}}\sum_{j=0}^{m^K-1}{\gamma_j[t]}\log{\PP(\rv[t],g^{-1}(\wv[t])=j|\thetav_j})\nonumber\\
&\;\;\;\;\;={\sum_{t=1}^{T_t+T_u}}\sum_{j=0}^{m^K-1}{\gamma_j[t]}(\log{p(\rv[t]|j,\thetav_j)}-K\log m),\label{eq:psi_def}
\end{align} where the second equality is from the Bayes rule and (\ref{eq:BLmodel}). Note that  $\gamma_j[t]$ in the above is constant with respect to $\thetav_j$. Also, from the Bernoulli-like model in (\ref{eq:BLmodel}), the objective function in (\ref{eq:psi_def}) can be specified as
\begin{align*}
\psi(\thetav|\thetav^i)&={\sum_{t=1}^{T_t+T_u}} {\sum_{j=0}^{m^K-1}}-{\gamma_{j}[t]}{K\log{m}}\nonumber\\
&+{\sum_{j=0}^{m^K-1}}{\sum_{t=1}^{T_t+T_u}}{\sum_{n=1}^{N}}\Big(\gamma_{j}[t]{{\bf 1}_{\{r_n{[t]} \neq c_{j,n} \}}}\log{\epsilon_{ j,n}}\nonumber\\
&\;\;\;\;\;\;\;\;\;\;\;\;\;\;\;\;\;\;\;\;\;\;\;\;+\gamma_{j}[t] {{\bf 1}_{\{r_n{[t]}=c_{j,n}\}}}{\log{(1-\epsilon_{j,n})}}\Big).
\end{align*} Since the first-term in the above is constant with respect to $\thetav$, the parameter vector $\thetav$ can be optimized by only maximizing the second-term as follows:
\begin{align}\label{eq:lttrterm}
&(\hat{\epsilonv}^{i+1},\hat{\cv}^{i+1})\nonumber\\
&=\argmax_{(\epsilonv,\cv)}{\sum_{j=0}^{m^K-1}}{\sum_{n=1}^{N}}{\sum_{t=1}^{T_t+T_u}}\Big({\gamma_{j}[t]}{{\bf 1}_{\{r_n{[t]} \neq c_{j,n} \}}}{\log{\epsilon_{j, n}}}\nonumber\\
&\;\;\;\;\;\;\;\;\;\;\;\;\;\;\;\;\;\;\;\;\;\;+{\gamma_{j}[t]}{{\bf 1}_{\{r_n{[t]}=c_{j, n}\}}}\log{(1-\epsilon_{j, n})}\Big).
\end{align} 
Obviously, we can see that maximizing \eqref{eq:lttrterm} is equivalent to maximizing the individual terms in \eqref{eq:lttrterm}: For each fixed $j$ and $n$, we have
\begin{align}\label{eq:indivterm}
&(\hat{\epsilon}_{j,n}^{i+1},\hat{c}_{j,n}^{i+1})=\argmax_{(\epsilon_{j,n}, c_{j,n})}{\sum_{t=1}^{T_t+T_u}}\Big({\gamma_{j}[t]}{{\bf 1}_{\{r_n{[t]} \neq c_{j, n} \}}}{\log{\epsilon_{j, n}}}\nonumber\\
&\;\;\;\;\;\;\;\;\;\;\;\;\;\;\;\;\;\;\;\;\;\;\;\;\;+{\gamma_{j}[t]}{{\bf 1}_{\{r_e{[t]}=c_{j, n}\}}}\log{(1-\epsilon_{j, n})}\Big).
\end{align} 
To solve the above problem, we introduce the useful lemma in the below.
\begin{lemma}\label{lem_hj}
Suppose $a_{\ell}\geq 0$ for $1\leq{\ell}\leq{n}$, Then $\sum_{\ell=1}^{n}{a_{\ell}}\log{p_{\ell}}$ is maximized over all probability vectors $p=(p_1,\ldots,p_n)$ by $p_{\ell}=\frac{a_{\ell}}{\sum_{i=1}^{n}{a_{i}}}$. \hfill$\blacksquare$
\end{lemma}
\vspace{0.2cm}

First of all, we observe that the optimal $c_{j,n}$ should satisfy the following constraint for any $\epsilon_{j,n} < 0.5$:
\begin{equation}\label{eq:constraint}
{\sum_{t=1}^{T_t+T_u}}{\gamma_{j}[t]}{{\bf 1}_{\{r_n{[t]} \neq c_{j, n} \}}}<{\sum_{t=1}^{T_t+T_u}}{\gamma_{j}[t]}{{\bf 1}_{\{r_n{[t]}=c_{j, n}\}}}.
\end{equation} Also, we can  see that this constraint is satisfied by assigning 
\begin{itemize}
\item  $\hat{c}_{j,n}^{i+1}=1$ if $\sum_{t=1}^{T_t+T_u}{r_{n}{[t]}}{{\gamma_{j}}[t]}>0$;
\item $\hat{c}_{j,n}^{i+1}=-1$ if $\sum_{t=1}^{T_t+T_u}{r_{n}{[t]}}{{\gamma_{j}}[t]}<0$.
\end{itemize} Equivalently, we obtain that
\begin{align}\label{eq:update1}
\hat{c}_{j, n}^{i+1}=\sign\left(\sum_{t=1}^{T_t+T_u}{\gamma_{j}[t]}r_n[t]\right) \mbox{ for } n \in [N].
\end{align} Next, applying Lemma~\ref{lem_hj} in the below to \eqref{eq:indivterm}, the error-probability $\epsilon_{j,n}^{i+1}$ is optimized as
\begin{align}\label{eq:update2}
&\hat{\epsilon}_{j, n}^{i+1}=\nonumber\\
&\frac{\sum_{t=1}^{T_t+T_u}{\gamma_{j}[t]}{{\bf 1}_{\{r_n [t] \neq \hat{c}_{j,n}^{i+1} \}}}}{\sum_{t=1}^{T_t+T_u}{\gamma_{j}[t]}{{\bf 1}_{\{r_n[t] \neq \hat{c}_{j, n}^{i+1} \}}}+\sum_{t=1}^{T_t+T_u}{\gamma_{j}[t]}{{\bf 1}_{\{r_n [t] \neq \hat{c}_{j,n}^{i+1} \}}}}.
\end{align} Finally, we can compute the log-likelihood \eqref{eq:likelihood} using the updated parameter vector $\thetav^{i+1}$ as
\begin{align}\label{eq:llikelihood}
\log{\PP(\Dc|\thetav^{i+1})}&={\sum_{t=1}^{T_t}}{\log{\frac{1}{m^K}p(\rv[t]|j_{t} ,\thetav_{j_{t} }^{i+1})}}\nonumber\\
&+\sum_{t=T_t+1}^{T_t+T_u}\log{\frac{1}{m^K}}{\sum_{j=0}^{m^K-1}{p(\rv[t]|j,\thetav_{j}^{i+1})}},
\end{align}  which is used to check the convergence of EM algorithm. The overall procedures are summarized in Fig. \ref{SSL concept} and {\bf Algorithm 1} where $\varepsilon \geq 0$ denotes the pre-determined threshold for the stopping criterion. 

{\bf Data Detection:} For $t\in[T_t+1:T_t+T_u]$, the SSL detector performs using the latest $\gamma_{j}[t]$ in \eqref{eq:expect}  as
\begin{equation}
\hat{j}=\argmax_{j\in [0:m^K-1]}{\gamma_{j}[t]}.
\label{eq:detects} 
\end{equation} 
Also, for $t\in[T_t+T_u+1:T_c]$, the detection process of the SSL detector is equivalent to that of  the SL detector in Section~\ref{sec:Supervised_Learning}. We remark that the performance-complexity tradeoff of the proposed SSL detector is controlled by the choice of $T_u$.



 
\begin{figure}[t]
\centerline{\includegraphics[width=9.5cm]{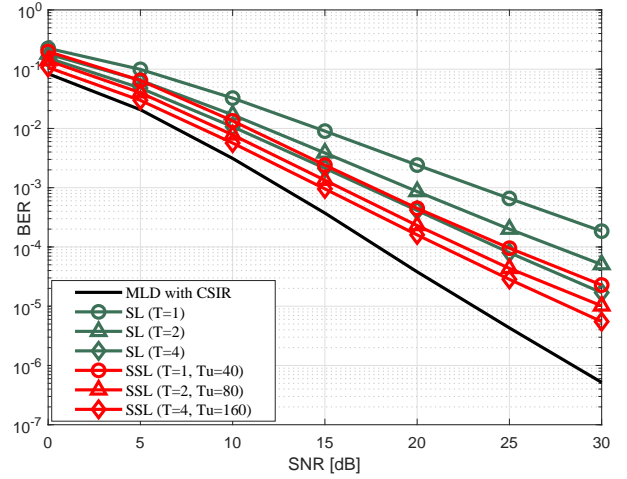}}\vspace{-0.3cm}
\caption{$K=2$ and $N=8$. Performance comparisons of the proposed SSL detector, the SL detector, and MLD with CSIR.}
\label{comp1}
\end{figure}

\begin{algorithm}\label{mbdetector}
\caption{Parameter update of the proposed SSL detector}
\textbf{Input:} 
\begin{itemize}
\item {\em (Labeled data)} $\Lc=\{(\rv[t],j_t): t\in[T_t]\}$
\item {\em (Unlabeled data)} $\Uc=\{\rv[t]: t\in [T_t+1:T_t+T_u]\}$
\end{itemize}
\textbf{Output:} $\hat{\thetav}^{i+1}$
\begin{algorithmic}
\State {Estimate ${\thetav}^0$ from $ \Lc$ using \eqref{eq:opt_mu} and \eqref{eq:est_epsilon}}
\State{Calculate log likelihood $\log{\PP(\Dc|\thetav^0)}$ from (\ref{eq:llikelihood})}
\While{$\log{\PP(\Dc|\thetav^{i+1})}-\log{\PP(\Dc|\thetav^i)} < \varepsilon $}
\For{$j=0,\ldots,m^K-1$}
\State {\;\;{\bf E-step:} Update $\{\gamma_{j}[t]: t\in [T_t+1:T_t+T_u]\}$ by $~~~~~~~~~~~~~~~~~~~~~$\eqref{eq:expect}}
\State {\;\;{\bf M-step:} Update $\thetav_j^{i+1}$ by \eqref{eq:update1} and \eqref{eq:update2}}
\EndFor
\State {Calculate log likelihood $\log{\PP(\Dc |\thetav^{i+1})}$ from (\ref{eq:llikelihood})} 
\State {Set $i=i+1$}
\EndWhile
\end{algorithmic}
\end{algorithm}

\section{Simulation Results}\label{simulation}

We evaluate the average bit-error rate (BER) performances of the proposed SSL detector and the conventional SL detector. For the simulations, a Rayleigh fading channel is considered where each element of a channel matrix {\bf H} is drawn from an independent and identically distributed (i.i.d.) circularly symmetric complex gaussian random variable with zero mean and unit variance. a user is assumed to send binary data ($m=2$) and QPSK modulation is applied. A block fading duration (i.e., coherence time interval) is set to be $T_d=512, T_u=10\cdot T_t$ and $T_t=T\cdot{m^K}$.

Fig.~\ref{comp1} shows the BER performances of the SSL detector, SL detector, and maximum likelihood detection (MLD) with channel state information at a receiver (CSIR) in a condition of various training duration. It is notable that the performance of proposed SSL detector outperforms the conventional SL detector in the entire SNR regimes where the pilot-overhead is same. In particular, for $T=1$, the performance of the proposed SSL detector almost achieves that of the SL detector with $T=4$. This implies that the SSL detector reduces training span ($T_t$) considerably without degradation in performance, by making the best use of information from the generative model and data signals. Also, when compared with MLD in CSIR, this result shows that the proposed method allows the empirical conditional probability to converge into true conditional probability without increasing the number of pilots.


\section{Conclusion}\label{conclusion}

In this paper, we presented a novel semi-supervised learning detector inspired by semi-supervised learning. Specifically, the proposed SSL detector updates parameters by using data signals through the maximum likelihood estimation under the Bernoulli-like model. Such parameter updates can significantly reduce pilot-overhead that is an issue in the existing SL detector. The simulation results demonstrated that the performance of the SSL detector almost achieves that of the SL detector, even with a quite lower pilot-overhead than that of the SL detector. We would like to emphasize that a SSL detector would be a strong practical framework in a field of machine learning based detector, in that compared with pilot signals, data signals are fairly cheap to obtain. On going work, we are investigating to develop more practical SSL detectors which require low complexity or are appropriate for time-varying channel system.


\section*{Acknowledgement}

This work was supported by Samsung Research Funding \& Incubation Center of Samsung Electronics under Project Number SRFC-IT1702-00.

\bibliographystyle{IEEEtran}
\bibliography{ICC_ref}%

\end{document}